\documentclass[12pt,twoside,]{pinp}

\usepackage[T1]{fontenc}
\usepackage[utf8]{inputenc}

\definecolor{pinpblue}{HTML}{185FAF}  % imagecolorpicker on blue for new R logo
\definecolor{pnasbluetext}{RGB}{101,0,0} %

\title{Parallel Computing With R: A Brief Review}

\author[1]{Dirk Eddelbuettel}

  \affil[1]{University of Illinois, Urbana-Champaign, IL, USA}

\leadauthor{Eddelbuettel}


\begin{abstract}
Parallel computing has established itself as another standard method for
applied research and data analysis. The R system, being internally
constrained to mostly singly-threaded operations, can nevertheless be
used along with different parallel computing approaches. This brief
review covers OpenMP and Intel TBB at the CPU- and compiler level, moves
to process-parallel approaches before discussing message-passing
parallelism and big data technologies for parallel processing such as
Spark, Docker and Kubernetes before concluding with a focus on the
future package integrating many of these approaches.
\end{abstract}

\dates{This version was compiled on \today} 

\doifooter{\url{https://arxiv.org/abs/1912.11144}}

\pinpfootercontents{Parallel Computing With R: A Brief Review}

\begin{document}

% Optional adjustment to line up main text (after abstract) of first page with line numbers, when using both lineno and twocolumn options.
% You should only change this length when you've finalised the article contents.
\verticaladjustment{-2pt}

\maketitle
\thispagestyle{firststyle}
\ifthenelse{\boolean{shortarticle}}{\ifthenelse{\boolean{singlecolumn}}{\abscontentformatted}{\abscontent}}{}

% If your first paragraph (i.e. with the \dropcap) contains a list environment (quote, quotation, theorem, definition, enumerate, itemize...), the line after the list may have some extra indentation. If this is the case, add \parshape=0 to the end of the list environment.

\hypertarget{introduction}{%
\section{Introduction}\label{introduction}}

A decade ago, \citet{SchmidbergerEtAl:2009:JSS} surveyed parallel
computing with the R language and environment. Their overview article
provided a useful road map for several years to come, and aided both
researchers and applied statisticians in making sense of a rapidly
changing landscape.

Roughly a decade later, a lot has changed. Entirely new technologies
such as Docker have emerged (see \citet{RJournal:Rocker} for a number of
use cases in the context of R). Cloud computing, which was just
beginning to make its case when the earlier article was written, is now
a dominant use case. Standard high-performance computing, a focus of the
earlier article, is still around but somewhat overshadowed by cloud
computing. New and more commercially-focused integrations such as Hadoop
and Spark offer a different take on data parallelism and have become
standard application in industry in part by their ability to cope with
`big(ger) data' requirements. At the same time, deep learning emerged as
a new field and brought with it several new new computing frameworks.
While a variety of competing frameworks are available, TensorFlow
\citep{Tensorflow-Whitepaper} and PyTorch \citep{NeurIPS:PyTorch} can be
considered as the two most prominent examples.

At the other end of the hardware spectrum, processing units offer more
and more parallelism directly at the center of the operations. Central
processing units (CPUs), still primarily provided by Intel and AMD, now
offer several dozen cores permitting both multi-threaded and
multi-process applications. Graphical processing units (GPUs) continue
to play a very important role as well by offering hundreds of cores
(albeit at a lower clock speed). Many high-end super-computing
installations (as \emph{e.g.} evidenced by the TOP500 listings) combine
CPUs and GPUs. A whole new entrant in the hardware space are dedicated
components for deep learning which are often data center-vendor
specific. One example is called Tensor Processing Units (TPUs) and
offered by Google, the company behind TensorFlow.

During this time, some parts remained the same. Compiler-supported
``local'' parallelism such as OpenMP has become fairly standard. It is
(within limits) used by R itself as well as in a number of application
packages as we will discuss in more detail below. But some new entrants
emerged as well. The future package by \citet{CRAN:future} offers an
elegant unified abstraction over both ``local'' and ``remote''
(\emph{i.e.}, distributed) variants, and has become popular and widely
adopted.

This paper updates the road map by \citet{SchmidbergerEtAl:2009:JSS} by
surveying several of these more recent entries. We begin by defining a
number of terms to facilitate the discussion. We then start from the
inside out, focusing first on OpenMP and related compiler-driven
technologies that are very `local' to the core of the execution. Next,
we consider more conventional high-performance and parallel computing
approaches. We then turn to a few of the newer frameworks which have
arisen in the last decade, and which can also play an important role in
parallel computing. Thereafter, we discuss the futures package as the
key application for R in this space. A brief summary concludes.

\hypertarget{definitions}{%
\section{Definitions}\label{definitions}}

Before discussing actual technologies and approaches for parallel
computing with R, it is helpful to review and clarify a number of terms
used in the discussion. This section recalls and illustrates a number of
standard definitions.

\hypertarget{sequential-execution}{%
\subsection{Sequential Execution}\label{sequential-execution}}

R as a language and environment is reasonably well established and
understood. A combination of dynamic typing, lazy functional evaluation
and object-orientation (in several flavors) makes for a somewhat unique
combination as discussed by \citet{MorandatEtAl:R}. One side-effect of
this design is that core operations are undertaken in
\emph{single-threaded} mode, or, in other words, sequentially.
\emph{Sequential computing} consists of operations executed in strict
sequence, and following a predetermined ordering. It may be the simplest
unit of computation (beyond a single instruction): a subroutine or
function in which each operation happens strictly after the preceding
one.

As an example and to fix the discussion on some concrete illustrations,
consider the following (simplified) code example. We omit error and
argument checking for the sake of brevity.

\begin{Shaded}
\begin{Highlighting}[]
\CommentTok{## apply func() to each element of vec, sum results}
\NormalTok{applyFuncToVec <-}\StringTok{ }\ControlFlowTok{function}\NormalTok{(vec, func) \{}
\NormalTok{    res <-}\StringTok{ }\DecValTok{0}
\NormalTok{    n <-}\StringTok{ }\KeywordTok{length}\NormalTok{(vec)}
    \ControlFlowTok{for}\NormalTok{ (i }\ControlFlowTok{in} \DecValTok{1}\OperatorTok{:}\NormalTok{n) \{}
\NormalTok{        res <-}\StringTok{ }\NormalTok{res }\OperatorTok{+}\StringTok{ }\KeywordTok{func}\NormalTok{(vec[i])}
\NormalTok{    \}}
\NormalTok{    res}
\NormalTok{\}}
\end{Highlighting}
\end{Shaded}

The principal computational cost of the function is likely the repeated
evaluation of the supplied function \texttt{func()} (provided its
complexity is sufficiently different from an ``empty'' function in which
case the overhead of the loop, and the repeated function calls
dominate). As should be clear from the code, all these operations are
executed sequentially with each of the \(n\) calls occurring after the
other.

Such a sequential setup provides the baseline against which parallel
execution can be measured in terms of both benefits (such as lower
execution time) and costs (such as increased memory or communication
overhead).

\hypertarget{concurrent-execution}{%
\subsection{Concurrent Execution}\label{concurrent-execution}}

The first generalization from sequential execution is \emph{concurrent}
execution. Concurrency is commonly defined as running more than one task
in non-overlapping time periods. In other words, at any one time only
\emph{one} task is executed however processing switches (possibly
frequently) between multiple tasks. A classic example is a computer
operating system advancing multiple tasks on a single CPU and single
core system: tasks are executed in turns, and possibly in small
increments. This can give the illusion of several things happening at
once; yet in reality this really is merely efficiently switching between
tasks.

Hence, a key aspects of concurrency is the \emph{task-switching cost}
which, at a minimum, involves memory access and moves to bring the
different tasks to the CPUs.

\hypertarget{parallel-execution}{%
\subsection{Parallel Execution}\label{parallel-execution}}

The polar opposite to sequential execution is \emph{parallel} execution.
It is defined as multiple operations happening in overlapping time
periods. This requires multiple execution units: maybe \emph{cores}
within a CPU, maybe multiple CPUs (possibly each with multiple cores),
and maybe multiple computers systems.

An important distinction hinges on the actual task being executed in
parallel, and its properties. A key class of problems are known as
\emph{data parallel} problems. Each sub-task \(i\) is independent of
each other sub-task \(j\). In such a case, no constraints are imposed on
the task scheduling. In fact, \emph{linear speedups} would be possible
were it not for the actual overhead in scheduling and coordinating
between the parallel workloads being executed. A classic example is our
simple function call above when the supplied function \texttt{func()}
depends only on the current vector element. Now multiple chunks of the
vector can be operated on in parallel, and the final result can be
aggregated. (Without going into finer details, this is the core
principle behind `map-reduce' computing approaches.)

Of course, the opposite also holds: should the tasks have \emph{data
dependence}, then parallel (or even concurrent) execution is much more
challenging. A simple example would be an instance of \texttt{func()}
where the \(i\)-th value depends on the preceding value with index
\(i-1\).

\hypertarget{single-instruction-multiple-data-simd}{%
\subsection{Single Instruction Multiple Data
(SIMD)}\label{single-instruction-multiple-data-simd}}

Another class of CPU and compiler-centric parallel instructions is
\emph{Single Instruction Multiple Data}, or SIMD. A good overview and
history is offered by \citet{Wikipedia:SIMD}.

The history of SIMD goes back to when `vector computers' were the
original supercomputers. However, such instruction sets are now also
common in high-end workstations and servers. The current top-of-the line
offering is \emph{AVX-512} instruction sets which operate on 512 bits at
a time. With standard double-precision floating point variables taking
eight bytes, or 64 bits, we see eight of these being addressable via
AVX-512 instruction sets. It should be noted that only higher-end server
CPUs and chip sets currently offer this. For example, the author's
workstation with a more modest Intel Core i7 processor, albeit at six
cores, does not offer it.

As such techniques are very dependent on the actual hardware used, and
vary with the hardware, benchmarking toolkits are important as they can
help in comparing across setups, and in calibration. One example of such
an (open source) toolkit is Likwid \citep{Likwid2010,Likwid2014}.

\hypertarget{local-parallelism-via-shared-memory-architecture}{%
\section{Local Parallelism via Shared Memory
Architecture}\label{local-parallelism-via-shared-memory-architecture}}

Parallel computing that is accessible from R can be implemented via
compiled extensions to R by relying on compiler-specific extensions.
This section discusses two: OpenMP, an industry standard for
shared-memory parallelism, and Intel Thread Building Blocks, which is
similar to OpenMP but offers a higher abstraction level.

\hypertarget{openmp}{%
\subsection{OpenMP}\label{openmp}}

A key technology for parallel execution of compiled code is provided by
OpenMP which was introduced by \citet{Dagum:1998}. It has become an
industry standard with the current version being OpenMP 5.0 published in
November 2018. Support for OpenMP is generally provided by the compiler.
For the GNU Compiler Collection, all of \texttt{gcc}, \texttt{g++} and
\texttt{gfortran} can use OpenMP directives. R itself has had support on
all platforms since release 3.4.0 in April 2017.

As mentioned earlier, and because of the dynamic nature of R, care has
to be taken in order to use OpenMP as R itself is single-threaded. That
said, OpenMP has long been used by R itself (conditional of course on
the particular build and architecture supporting it). The use of OpenMP
ranges from an option for matrix multiplication still labeled
``experimental'' (see the `matprod' entry in \texttt{help(options)}) to
possible OpenMP/SIMD enhanced checks for non-finite values as well as
parallel column sums (all in \texttt{src/main/array.c}) and to the use
in the implementation of the \texttt{dist()} function in the base R
package \texttt{stats}.

\begin{Shaded}
\begin{Highlighting}[]
\AttributeTok{static}\NormalTok{ Rboolean mayHaveNaNOrInf_simd(}\DataTypeTok{double}\NormalTok{ *x, }\DataTypeTok{R_xlen_t}\NormalTok{ n)}
\NormalTok{\{}
  \DataTypeTok{double}\NormalTok{ s = }\DecValTok{0}\NormalTok{;}
  \CommentTok{/* SIMD reduction is supported since OpenMP 4.0. The value of _OPENMP is}
\CommentTok{     unreliable in some compilers, so we depend on HAVE_OPENMP_SIMDRED,}
\CommentTok{     which is normally set by configure based on a test. */}
  \CommentTok{/* _OPENMP >= 201307 */}
\PreprocessorTok{#if defined(_OPENMP) && HAVE_OPENMP_SIMDRED}
  \PreprocessorTok{#pragma omp simd reduction(+:s)}
\PreprocessorTok{#endif}
  \ControlFlowTok{for}\NormalTok{ (}\DataTypeTok{R_xlen_t}\NormalTok{ i = }\DecValTok{0}\NormalTok{; i < n; i++)}
\NormalTok{    s += x[i];}
  \ControlFlowTok{return}\NormalTok{ !R_FINITE(s);}
\NormalTok{\}}
\end{Highlighting}
\end{Shaded}

This is a typical usage pattern for OpenMP. The C code is ornamented
with a \texttt{\#pragma} which is itself conditional on OpenMP as well
as SIMD reductions with OpenMP being available. If so the compiler will
generate more efficient parallel code: \emph{in lieu} of the basic loop
touching each vector element in sequence, SIMD operations can do so with
some parallelism.

As noted above, SIMD support is also dependent on the particular CPU
model. So for a concrete example (which is shown below), we compare a
simpler OpenMP \texttt{parallel\ for} loop with a normal,
non-parallelized loop. It can be built via
\texttt{Rcpp::sourceCpp("filename.cpp")}, and relies on the Rcpp
\citep{CRAN:Rcpp,TAS:Rcpp} and microbenchmark
\citep{CRAN:microbenchmark} packages. The benchmarking section at its
end will run when the code is sourced.

\begin{Shaded}
\begin{Highlighting}[]
\PreprocessorTok{#include }\ImportTok{<Rcpp.h>}

\CommentTok{// [[Rcpp::plugins(openmp)]]}

\CommentTok{// [[Rcpp::export]]}
\DataTypeTok{bool}\NormalTok{ mayHaveNaNOrInf_openmp(Rcpp::NumericVector x) \{}
  \DataTypeTok{size_t}\NormalTok{ n = x.size();}
  \DataTypeTok{double}\NormalTok{ s = }\DecValTok{0}\NormalTok{;}
\PreprocessorTok{#if defined(_OPENMP)}
  \PreprocessorTok{#pragma omp parallel for}
\PreprocessorTok{#endif}
  \ControlFlowTok{for}\NormalTok{ (}\DataTypeTok{size_t}\NormalTok{ i = }\DecValTok{0}\NormalTok{; i < n; i++)}
\NormalTok{    s += x[i];}
  \ControlFlowTok{return}\NormalTok{ !R_FINITE(s);}
\NormalTok{\}}

\CommentTok{// [[Rcpp::export]]}
\DataTypeTok{bool}\NormalTok{ mayHaveNaNOrInf(Rcpp::NumericVector x) \{}
  \DataTypeTok{size_t}\NormalTok{ n = x.size();}
  \DataTypeTok{double}\NormalTok{ s = }\DecValTok{0}\NormalTok{;}
  \ControlFlowTok{for}\NormalTok{ (}\DataTypeTok{size_t}\NormalTok{ i = }\DecValTok{0}\NormalTok{; i < n; i++)}
\NormalTok{    s += x[i];}
  \ControlFlowTok{return}\NormalTok{ !R_FINITE(s);}
\NormalTok{\}}

\CommentTok{/*** R}
\CommentTok{library(microbenchmark)}
\CommentTok{x <- rep(0, times=1e7)}
\CommentTok{Sys.setenv("OMP_NUM_THREADS"=6)}
\CommentTok{options(digits=3) # more compact}
\CommentTok{microbenchmark(mayHaveNaNOrInf_openmp(x),}
\CommentTok{               mayHaveNaNOrInf(x), times=1000)}
\CommentTok{*/}
\end{Highlighting}
\end{Shaded}

On an Intel Core i7-8700 with six cores, we obtain a roughly four-fold
improvement (when measuring median time) between the conventional loop,
and the OpenMP-parallelized loop. That is a significant speed gain for
an operation that may be executed quite frequently, and shows why OpenMP
is a compelling tool. However, it also shows that OpenMP requires some
familiarity with programming at the source level (in C, C++, or
Fortran).

OpenMP is a key technology for parallel computing with R. How to enable
it is described in some detail in Section ``OpenMP support'' of the
``Writing R Extensions'' manual \citep{CRAN:R}. We should also note that
different operating system vary somewhat in how well they support
OpenMP. Since the R 3.4.0 release in 2017, it is generally available
across the platforms supported by R. The rich corpus of package source
code presented by the CRAN repositories offers some insights in how
prevalent the usage of OpenMP among R packages is. We find 211 packages
using OpenMP via the R Core team recommended use of common \texttt{make}
variables set by R during its build. Here,
\texttt{SHLIB\_OPENMP\_CFLAGS} is the C compiler flags, similar flags
exist for C++ as well as for Fortran. Running a simple \texttt{grep}
among the \texttt{Makevars\{,.in\}} files reveals 355 packages using
this instruction. This can be seen as an upper bound as some packages
recommend use of the build variable in order to benefit from OpenMP use
in the code they provide for other packages. An example of this practice
is RcppArmadillo \citep{CRAN:RcppArmadillo}) offering parallel linear
algebra and utility code for use by other packages.

Going one step further, we can look for actual \texttt{\#pragma\ omp}
code in the package sources. We consider code in or below the
\texttt{src/} directory, and also in or below the \texttt{inst/include}
directory from where it may be accessible to other packages via header
includes. We find that over two-hundred R packages on CRAN use OpenMP.
Among these, StanHeaders \citep{CRAN:StanHeaders} leads with 246
instances, the RViennaCL package \citep{CRAN:RViennaCL} has 232, spMC
\citep{CRAN:spMC} is not far behind with 226 and packages data.table
\citep{CRAN:data.table}, xgboost \citep{CRAN:xgboost}, and RNiftyReg
\citep{CRAN:RNiftyReg} each have over 90. This informal inspection of
the CRAN package sources shows that OpenMP is indeed widely-used among R
packages.

\hypertarget{intel-thread-building-blocks-tbb}{%
\subsection{Intel Thread Building Blocks
(TBB)}\label{intel-thread-building-blocks-tbb}}

A closely-related technology is the Intel Thread Building Blocks (or
Intel TBB for short). Like OpenMP, it is a means to have the compiler
generate parallel code but in this case at a somewhat higher abstraction
level, and only via C++.

The RcppParallel package \citep{CRAN:RcppParallel} wraps the Intel TBB
making it straightforward to deploy this technology, especially if R and
C++ bindings are already being used. The package comes with
documentation and examples showing typical Map-Reduce patterns where the
Intel TBB can automatically determine the optimal `chunk' sizes when
deciding how to parallelise, say, operations on a vector just as in our
first example above.

The following code example, taken from the RcppParallel documentation,
illustrates how a worker object for a `summation' can be set up. The
example is similar to our initial example as it operates element-wise on
a vector. Here, a key element is in the \texttt{operator()}
implementation: by giving \texttt{begin} and \texttt{end} indices, we
allow the worker to operator on the contiguous chunk of the vector
defined by this index range. One level up, TBB deals with appropriate
`chunking' and submits each `chunk' to the worker shown here. A
\texttt{join()} operation updates an incoming \texttt{Sum} variable by
adding the results of the computations on just this chunk.

\begin{Shaded}
\begin{Highlighting}[]
\CommentTok{// [[Rcpp::depends(RcppParallel)]]}
\PreprocessorTok{#include }\ImportTok{<RcppParallel.h>}
\KeywordTok{using} \KeywordTok{namespace}\NormalTok{ RcppParallel;}

\KeywordTok{struct}\NormalTok{ Sum : }\KeywordTok{public}\NormalTok{ Worker \{   }
  \CommentTok{// source vector}
  \AttributeTok{const}\NormalTok{ RVector<}\DataTypeTok{double}\NormalTok{> input;}
   
  \CommentTok{// accumulated value}
  \DataTypeTok{double}\NormalTok{ value;}
   
  \CommentTok{// constructors}
\NormalTok{  Sum(}\AttributeTok{const}\NormalTok{ NumericVector input) : input(input), value(}\DecValTok{0}\NormalTok{) \{\}}
\NormalTok{  Sum(}\AttributeTok{const}\NormalTok{ Sum& sum, Split) : input(sum.input), value(}\DecValTok{0}\NormalTok{) \{\}}
   
  \CommentTok{// accumulate just the element of the range I've been asked to}
  \DataTypeTok{void} \KeywordTok{operator}\NormalTok{()(}\BuiltInTok{std::}\NormalTok{size_t begin, }\BuiltInTok{std::}\NormalTok{size_t end) \{}
\NormalTok{     value += }\BuiltInTok{std::}\NormalTok{accumulate(input.begin()+begin, input.begin()+end, }\FloatTok{0.0}\NormalTok{);}
\NormalTok{  \}}
     
  \CommentTok{// join my value with that of another Sum}
  \DataTypeTok{void}\NormalTok{ join(}\AttributeTok{const}\NormalTok{ Sum& rhs) \{ }
\NormalTok{     value += rhs.value; }
\NormalTok{  \}}
\NormalTok{\};}
\end{Highlighting}
\end{Shaded}

The attractive high-level abstraction of Intel TBB is clearly apparent
in the following example. We combine a \texttt{Sum} instance (utilizing
the \texttt{Worker} shown in the previous example) with a given numeric
vector \texttt{x}, and pass it to \texttt{parallelReduce()} along with
the dimensions (here the entire vector length). Intel TBB takes care of
appropriate task and chunk size as well as scheduling, even offering
what is referred to as `work-stealing' where idle cores can take up
processing from other queues when their own queue is empty. But these
details are hidden by the implementation---we simple return the computed
(scalar) value at the end.

\begin{Shaded}
\begin{Highlighting}[]
\CommentTok{// [[Rcpp::export]]}
\DataTypeTok{double}\NormalTok{ parallelVectorSum(NumericVector x) \{}
   
  \CommentTok{// declare the Sum instance }
\NormalTok{  Sum sum(x);}
   
  \CommentTok{// call parallel_reduce to start the work}
\NormalTok{  parallelReduce(}\DecValTok{0}\NormalTok{, x.length(), sum);}
   
  \CommentTok{// return the computed sum}
  \ControlFlowTok{return}\NormalTok{ sum.value;}
\NormalTok{\}}
\end{Highlighting}
\end{Shaded}

Intel TBB is an attractive solution if the C++ language can be used. It
is also not restricted to hardware manufactured by Intel as AMD CPUs
work as well. For R users, the RcppParallel package
\citep{CRAN:RcppParallel} makes it accessible from R by supplying all
required components, notably the Intel TBB header library itself, in the
convenient form of a package.

\hypertarget{other-related-aspects}{%
\subsection{Other Related Aspects}\label{other-related-aspects}}

One important topic related to both OpenMP and Intel TBB is the use of
random number generators (RNGs). For sequential code, an RNG
implementation can remain stateful and deliver high-quality random
draws. For a user of R, choosing one of the built-in RNGs that ship with
R is often a reasonable choice. The situation changes when it comes to
parallel execution. Here, we need stream-aware random-number creation
which is provided by both the older packages rlecuyer
\citep{CRAN:rlecuyer} and rstream \citep{CRAN:rstream} as well as the
more-recent ones dqrng \citep{CRAN:dqrng} and sitmo \citep{CRAN:sitmo}.

Another technology worth mentioning is ArrayFire, provided for R by the
(GitHub-only) package RcppArrayFire \citep{GH:RcppArrayFire}. It offers
a C++ abstraction to highly-parallel code which can be executed in three
different forms. The first is via multi-threading and OpenMP utilizing
only the CPU. This provides a base case usable on most computers. But
ArrayFire also provides wrappers to GPU code deploying either to CUDA
and OpenCL, if available.

Finally, we mention package RhpcBLASctl \citep{CRAN:RhpcBLASctl} which
provides a simple interface to querying and selecting the number of
threads used for the OpenBLAS numerical analysis library but also for
direct use of OpenMP (as it uses the same underlying environment
variable).

\hypertarget{machine-local-parallelism-with-many-cores}{%
\section{Machine-local Parallelism with Many
Cores}\label{machine-local-parallelism-with-many-cores}}

The previous section focused on a widely-available and widely-used
technology. Due to its dependence on a C, C++ or Fortran compiler, it
may be out of reach for some users. This section describes an
alternative which can be seen as the opposite in terms of accessibility:
using multiple cores on a single machine directly from R.

Nowadays, machines with `many cores' span the spectrum from servers to
workstations and laptops. At the upper end, servers running the Linux
operating system (or even Windows if one prefers or requires it) can
have several dozen cores. The largest `R5' instance on Amazon AWS is
listed (in December 2019) as having 96 cores and 768 GB of memory. Base
R, as shipped, can use this directly via the parallel package to deploy
all of these 96 cores for (process-parallel) work. The parallel package
was added to base R 2.14.0 (released in October 2011) after
incorporating key components from the multicore and snow packages
(already described by \citet{SchmidbergerEtAl:2009:JSS}). Both of these
packages can be considered deprecated in favor of the parallel package.

In the simplest use case of package parallel, one replaces
\texttt{lapply()} and \texttt{mapply()} with the corresponding
``multicore'' \texttt{mclapply()} and \texttt{mcmapply()} functions. The
following example run an `existence' proof requesting the process id of
the running R process. It invokes \texttt{mclapply()} with argument
\texttt{1:4} where a function of choice (here a simple request of the
process id) is called in parallel on these elements. Because four
elements and four cores (\texttt{mc.cores\ =\ 4}) are used, the result
of this call will be four unique process ids.

\begin{Shaded}
\begin{Highlighting}[]
\NormalTok{parallel}\OperatorTok{::}\KeywordTok{mclapply}\NormalTok{(}\DecValTok{1}\OperatorTok{:}\DecValTok{4}\NormalTok{, }\ControlFlowTok{function}\NormalTok{(i) }\KeywordTok{Sys.getpid}\NormalTok{(), }\DataTypeTok{mc.cores =} \DecValTok{4}\NormalTok{)}
\end{Highlighting}
\end{Shaded}

A major advantage of the multicore processing in parallel due to
\emph{forked} parallel processing is that global variables in the main R
session are inherited by the child processes. This means the developer
does not have to spend efforts on identifying and exporting those to the
parallel workers. However, one main disadvantage of forked parallel
processes in combination with low-level multi-threaded processing is
potential instability that may result in crashes and/or corrupt results.
Because of this, multicore parallelization can only safely be used in
cases where one has full control of all dependencies. If not, there is a
risk that the implemented pipeline breaks because a previously
single-threaded package dependency is updated to use multi-threaded
processing. Another disadvantage of multicore processing is that it is
only supported on Unix-like operating systems; on Microsoft Windows the
type of forked processing needed is unsupported and therefore, if used,
\texttt{mclapply()} and friends are designed to fall back to run in
sequential mode.

An alternative to \emph{multicore} processing is \emph{cluster}
processing, where parallelization takes place over a set of independent,
non-interactive R processes running in the background and waiting for
instructions from the main R session. The parallel package provides
functions to create such clusters and \texttt{*apply()}-like functions
to invoke a function in parallel across a cluster,
e.g.~\texttt{parLapply()} and \texttt{parMapply()}.

\begin{Shaded}
\begin{Highlighting}[]
\NormalTok{workers <-}\StringTok{ }\NormalTok{parallel}\OperatorTok{::}\KeywordTok{makeCluster}\NormalTok{(}\DecValTok{4}\NormalTok{, }\DataTypeTok{type =} \StringTok{"PSOCK"}\NormalTok{)}
\NormalTok{parallel}\OperatorTok{::}\KeywordTok{parLapply}\NormalTok{(}\DecValTok{1}\OperatorTok{:}\DecValTok{4}\NormalTok{, }\ControlFlowTok{function}\NormalTok{(i) }\KeywordTok{Sys.getpid}\NormalTok{(), }\DataTypeTok{cl =}\NormalTok{ workers)}
\end{Highlighting}
\end{Shaded}

The advantages of this model is that it is supported on all operating
systems, and that it also works with code that runs multi-threaded
internally. The disadvantages are increased communication overhead, and
that global variables have to be identified and explicitly exported to
each worker in the cluster before processing. As discussed below,
another advantage with cluster processing is that it supports also
workers on external machines, possibly running in remote locations.

Similarly, other packages besides parallel can be deployed. We will
discuss the future package \citep{CRAN:future} in more detail below, but
one simple use case for it is to replace the parallel package. Packages
snow \citep{CRAN:snow} (already described by
\citet{SchmidbergerEtAl:2009:JSS}), snowfall \citep{CRAN:snowfall} (a
once-popular ``simpler'' interface to snow) as well as foreach
\citep{CRAN:foreach} with its companion packages doMC \citep{CRAN:doMC}
and doParallel \citep{CRAN:doParallel} also offer alternatives for such
machine-local parallel tasks (as well as opportunities to launch
multi-machine computations as discussed in the next section) but are
seeing less active development than future.

\hypertarget{local-to-remote-parallelism-with-message-passing}{%
\section{Local to Remote Parallelism with Message
Passing}\label{local-to-remote-parallelism-with-message-passing}}

The Message Passing Interface (MPI) standard, first described in
\citet{MPI:Standard}, has long been a common tool in parallel and
high-performance computing. The OpenMPI implementation of the MPI
standard has now established itself as the reference implementation. It
is being coordinated by an industry consortium just like the OpenMP
standard discussed in the previous section. (Nowadays, the alternate
MPICH implementation is less relevant, as is the competing PVM
standard.)

In general, using MPI requires i) compiling and linking with a dedicated
MPI library such as OpenMPI, and ii) running the binary in a setup with
appropriate configuration. Given the message-passing nature, there
(generally) needs to be other hosts ready to receive (and send) messages
and to share the load (though for debugging and testing machine-local
setups are of course also possible). At a minimum, authentication,
permissioning and logging have to be provided making this a technology
that is mostly provisioned by data centers. And, as above for OpenMP,
programming with MPI at the C, C++ or Fortran level is not trivial. A
quick scan of the CRAN package universe reveals eighteen packages
including the \texttt{mpi.h} header file in their sources.

However, for parallel applications with R, the higher-level Rmpi package
\citep{CRAN:Rmpi} helps as it shields the user from some of the
technicalities by offering simpler higher-level constructs. This was
already described by \citet{SchmidbergerEtAl:2009:JSS} and has, by and
large, not changed much. The packages snow \citep{CRAN:snow}, snowfall
\citep{CRAN:snowfall} and doMPI \citep{CRAN:doMPI} are also still
available and can be used for these tasks. As mentioned earlier, they
are less actively developed than future \citep{CRAN:future} discussed
below. When using snow, snowfall or doMPI, the underlying OpenMPI
implementation can be used as the communications strategy in the cluster
object, and functions such as \texttt{parLapply()} and
\texttt{parSapply()} can be used to dispatch jobs across the cluster.

Some additional R packages of interest have emerged. Parallel computing
workloads are often executed by schedulers and similar frameworks
dealing with permissioning, quotas, logging, and other details. Among
several of these schedulers and related tools, the Slurm workload
manager has become one of the more popular choices. It offers a wide
range of features, and is distributed under an open source license with
optional commercial support. For use from R, the rslurm package
\citep{CRAN:rslurm} offers suitable integration. Its key functions
comprise both an \texttt{apply()} variant named \texttt{slurm\_apply()}
as well as single job launchers and job status query runners. Also of
note is package batchtools \citep{CRAN:batchtools} which operates in the
same space and lets R user deploy tasks across (larger) computational
clusters frequently used with MPI.

\hypertarget{frameworks-for-parallel-workloads-and-big-data}{%
\section{Frameworks for Parallel Workloads and ``Big
Data''}\label{frameworks-for-parallel-workloads-and-big-data}}

The previous three sections discussed what we may describe as ``classic
approaches'' to parallel computing. These approaches were also already
described in some form in \citet{SchmidbergerEtAl:2009:JSS}. They also
have one common element: all of them originate from, or are associated
with, the more research-oriented or scientific side of computing that is
commonly referred to as `high-performance computing' or `scientific
computing'.

Focusing on the other end of the spectrum, this section now turns
towards the `internet-scale' age of `Big Data'. We will look at
frameworks for analysing large data sets via Hadoop and Spark, two
technologies that have been dominating this area since the report by
\citet{SchmidbergerEtAl:2009:JSS} (which did not foresee them) came out.
We will then turn to containers for development and deployment which can
also be used for parallel computing.

\hypertarget{hadoop-and-spark}{%
\subsection{Hadoop and Spark}\label{hadoop-and-spark}}

The Hadoop File System \citep{Shvachko:2010}, frequently referred to as
just `Hadoop', started as a re-implementation of two key Google
technologies: MapReduce, and the Google File System. Its key attraction
is the ``unlimited'' scope of data it can process as a decentralized
federation of machines can be assembled. Hadoop, and later Spark,
clusters can be used ``on premises'' (\emph{i.e.} in local data centers)
as well as in cloud-based deployments. An entire eco-system of related
services such as the Hadoop Distributed File System and the Yarn
resource manager / scheduler emerged over time. All of these components
eventually became part of the larger Apache Project. Hadoop may now be
considered well-established and is certainly widely used, especially in
industry, but is generally no longer considered to be cutting edge.

Spark \citep{Zaharia:Spark:2010,CACM:Spark}, also an Apache Project,
originated at the University of California, Berkeley. It offers
resilient distributed data structures organized as data.frame objects,
making it suitable for very large data sets. Access from R is possible
via several projects, notably SparkR \citep{CRAN:SparkR} and sparklyr
\citep{CRAN:sparklyr}. Here the first package goes back to the Apache
Spark project itself, whereas the second package is a RStudio product
and well integrated into their offerings. This combines the strength of
Spark and its ability to access large distributed data sets with the
familiar R front-ends and operations, especially the dplyr package
\citep{CRAN:dplyr} for which sparklyr can act as a back-end and
connection to Spark.

At the time of writing, Spark can be considered as the dominant ``big
data'' platform in industry. Essentially all data-storage and processing
back-ends as well as analytics libraries connect to it, and can compute
results within Spark-based workflows. Spark, just like Hadoop, has a
reasonably strong connection to the Java ecosystem, and many
integrations utilize the Java Virtual Machine abstraction. It is however
also a `wire protocol' so non-JVM approaches can be used with it as
well.

\hypertarget{docker-and-kubernetes}{%
\subsection{Docker and Kubernetes}\label{docker-and-kubernetes}}

Another important and recent trend not anticipated in
\citet{SchmidbergerEtAl:2009:JSS} is containerization. Probably
best-known, and sometimes taken as synonymous, is Docker
\citep{Merkel:Docker:2014}. At its core, Docker relies on Linux-specific
resource management and isolation capabilities such as kernel namespaces
and cgroups to allow independent entities (the ``containers'') to run
within a single Linux host instance while avoiding the heavier overhead
of full virtual machines. Docker can be used on other operating systems
by providing a minimal layer of code that provides these calls; in the
simplest cases a virtual machine may be used. Docker is one of several
technologies using these features, but has become `the public face' of
container use eclipsing related approaches such as the ``Linux
Containers'' technology LXC \citep{Wikipedia:LXC}, or the Singularity
approach \citep{PLOS:Singularity} coming from the high-performance
computing side.

Within the R world, the Rocker Project, described by
\citet{RJournal:Rocker}, offers a wide variety of R-based containers.
These range from smaller and lighter-weight containers providing just R
(as well as development and test version of R) to fuller-featured
containers with RStudio Server, Shiny Server, or some larger collections
of packages, for example for geo-spatial computing. Another popular
series of containers `snapshots' to the R releases ensuring
reproducibility for various ``points in time''. Containers from the
Rocker Project also form the building blocks for a large number of
products utilizing their standardized packaging and provisioning of the
R system and packages.

Containerization, due to some key architectural and implementation
aspects, generally wraps a single process. This is a key feature which
makes it easy to ``wrap'' a single program or application along with all
its required libraries or components, and a clear key to its wide
adoption and success. Yet another common use case may involve the need
to have multiple programs operate in a coordinated manner. Such
``orchestration'' proved to be a very natural second step for Docker,
and multiple approaches have been tried.

Among these, the Kubernetes system \citep{Wikipedia:K8s} for container
orchestration originally started by Google (and now developed by the
Cloud Native Computing Foundation) has become the dominant entry.
Kubernetes is generally supported by cloud-computing providers as a
basic platform or infrastructure building block. It is frequently
combined with Spark, which we discussed in the previous section, to
provision ``big data'' computing at scale.

\hypertarget{application-focus-future}{%
\section{Application Focus: future}\label{application-focus-future}}

The future package by \citet{CRAN:future} provides a very attractive and
cohesive application layer for several topics discussed above. It builds
upon the concepts of futures \citep{HewittBaker_1977} and promises
\citep{FriedmanWise_1978, Hibbard_1976}. With a single unified
application-programming interface (API), it can replace simple uses
cases such as the \texttt{mclapply()} example above. Yet at the same
time it can also scale to multi-machine or multi-host parallel computing
using a variety of parallel computing back-ends.

One key element is an underlying abstraction from asynchronous
computing: future allows to ``launch'' a compute task and access /
retrieve its result ``when it is ready'' (or block until it is).
Consider a base R expression such as \texttt{a\ \textless{}-\ expr}
where the expression part evaluating \texttt{expr} will typically
involve some computation. Using the API of the future package, we can
write \texttt{f\ \textless{}-\ future(expr)} which ``hands'' the
expression to the future package, and a (later, or immediate)
\texttt{v\ \textless{}-\ value(f)}. The state of this future value is
either resolved, or unresolved in which case further execution may block
until it is resolved.

As an illustration, consider the following example borrowed from the
overview vignette. Simply by switching the \texttt{plan()} between
sequential, multi-process and multi-machine, the well-known (and
time-consuming) Mandelbrot Set illustration can be rendered in three
different ways at three different speeds.

\begin{Shaded}
\begin{Highlighting}[]
\KeywordTok{library}\NormalTok{(}\StringTok{"future"}\NormalTok{)}

\CommentTok{# base case: sequantial local execution}
\KeywordTok{plan}\NormalTok{(sequential)                        }
\KeywordTok{demo}\NormalTok{(}\StringTok{"mandelbrot"}\NormalTok{, }\DataTypeTok{package =} \StringTok{"future"}\NormalTok{, }\DataTypeTok{ask =} \OtherTok{FALSE}\NormalTok{)}

\CommentTok{# multiprocess parallelizes}
\KeywordTok{plan}\NormalTok{(multiprocess)                      }
\KeywordTok{demo}\NormalTok{(}\StringTok{"mandelbrot"}\NormalTok{, }\DataTypeTok{package =} \StringTok{"future"}\NormalTok{, }\DataTypeTok{ask =} \OtherTok{FALSE}\NormalTok{)}

\CommentTok{# if work node machines 'n2' to 'n9' are available, use them}
\KeywordTok{plan}\NormalTok{(cluster, }\DataTypeTok{workers =} \KeywordTok{c}\NormalTok{(}\StringTok{"n2"}\NormalTok{, }\StringTok{"n5"}\NormalTok{, }\StringTok{"n6"}\NormalTok{, }\StringTok{"n6"}\NormalTok{, }\StringTok{"n9"}\NormalTok{))}
\KeywordTok{demo}\NormalTok{(}\StringTok{"mandelbrot"}\NormalTok{, }\DataTypeTok{package =} \StringTok{"future"}\NormalTok{, }\DataTypeTok{ask =} \OtherTok{FALSE}\NormalTok{)}
\end{Highlighting}
\end{Shaded}

The future package is complemented by several additional packages. The
future.apply package \citep{CRAN:future.apply} offers variants of R's
\texttt{*apply} functions, similar to the solutions discussed above but
also addressing load balancing aspects, that work with the different
back-ends supported by future. Back-ends are generally selected by
issueing a \texttt{plan()} statement with an appropriate processing
plan. This covers multi-process, cluster, as well as support for
computing-center scale via slurm, batchtools and more. In general, each
specific back-end is provided by an additional (custom) package.

Future offers a very consistent API. This facilitates adapting
deployment to different usage scenarios or backends. In many cases,
users will only have to change the \texttt{plan()} statement which
serves as the \emph{ex ante} declaration of the deployment. Further,
several detailed vignettes document usage of the future package, its
related packages, and its various options.

\hypertarget{summary}{%
\section{Summary}\label{summary}}

Parallel computing is a tool which can help in the context of large and
demanding analysis tasks, especially when these are seen to be
``data-parallel'' without interdependence or linkage between the tasks.
For the R user and analyst, parallel computing can take many forms. This
survey delineates the various possibilities from the inside out.
Starting with from `CPU-local' and compiler-dependendent methods like
OpenMP and Intel TBB, we looked at the machine-local multiprocessing
approaches, before discussing message-passing parallelism as offered by
OpenMPI. (Relatively) newer approaches centered on ``big data'' such as
Spark, Docker and Kubernetes rounded out the review along with an
application-level highlight offered by the future package.

It is our hope that this survey proves to be of use for analysts and
researchers wishing to analyse and model with R in a parallel manner.

\hypertarget{acknowledgements}{%
\section{Acknowledgements}\label{acknowledgements}}

This paper has benefited greatly from conversations with James Balamuta
and Henrik Bengtsson whose insightful suggestions are truly appreciated.
Further comments by the David Scott, the editor, and two anonymous
referees were also very helpful and are gratefully acknowledged.

%\showmatmethods

\bibliography{bibliography}
\bibliographystyle{jss}

\end{document}